\documentclass[12pt,preprint]{emulateapj}
\usepackage{epstopdf}

\shorttitle{Mass--size evolution of passive early--type galaxies at $0<z<3$}
\shortauthors{Cassata et al.}

\begin{document}

\title{Constraining The Assembly Of Normal And Compact Passively
  Evolving Galaxies From Redshift $z=3$ To The Present With CANDELS
  \altaffilmark{1}}

\altaffiltext{1}{paolo.cassata@oamp.fr}

\author{P. Cassata\altaffilmark{2}, 
M. Giavalisco\altaffilmark{3}, 
C. C. Williams\altaffilmark{3},
Yicheng Guo\altaffilmark{4},
Bomee Lee\altaffilmark{3},
A. Renzini\altaffilmark{5},
H. Ferguson\altaffilmark{6},
S. F. Faber\altaffilmark{4},
G. Barro\altaffilmark{4},
D. H. McIntosh\altaffilmark{7},
Yu Lu\altaffilmark{8},
E. F. Bell\altaffilmark{9},
D. C. Koo\altaffilmark{4},
C. J. Papovich\altaffilmark{10},
R. E. Ryan\altaffilmark{6},
C. J. Conselice\altaffilmark{11}
N. Grogin\altaffilmark{6},
A. Koekemoer\altaffilmark{6},
N. P. Hathi\altaffilmark{12}
}

\altaffiltext{2}{Aix Marseille Universit\'e, CNRS, LAM (Laboratoire
  d'Astrophysique de Marseille) UMR 7326, 13388, Marseille, France;
  paolo.cassata@oamp.fr} 
\altaffiltext{3}{Department of Astronomy,
  University of Massachusetts, Amherst, MA 01003}
\altaffiltext{4}{UCO/Lick Observatory, Department of Astronomy and
  Astrophysics, University of California, Santa Cruz, CA, USA}
\altaffiltext{5}{Osservatorio Astronomico di Padova (INAF--OAPD),
  Vicolo dell'Osservatorio 5, I-35122, Padova Italy}
\altaffiltext{6}{Space Telescope Science Institute, 3700 San Martin
  Boulevard, Baltimore, MD, 21218} 
\altaffiltext{7}{Department of Physics, University of Missouri, Kansas City, MO, USA}
\altaffiltext{8}{Kavli Institute for Particle Astrophysics and Cosmology, Stanford, CA 94309, USA}
\altaffiltext{9}{Department of Astronomy, University of Michigan, 500 Church Street, Ann Arbor, MI 48109, USA}
\altaffiltext{10}{Department of Physics and Astronomy, Texas A\&M University, College Station, TX, USA}
\altaffiltext{11}{The School of Physics and Astronomy, University of Nottingham, Nottingham, UK}
\altaffiltext{12}{Carnegie Observatories, Pasadena, CA 91101, USA}

\begin{abstract}
We study the evolution of the number density, as a function of the
size, of passive early--type galaxies with a wide range of stellar
masses ($10^{10} M_{\odot}<M_*\lesssim 10^{11.5} M_{\odot}$) from
$z\sim3$ to $z\sim1$, exploiting the unique dataset available in the
GOODS--South field, including the recently obtained WFC3 images as a
part of the Cosmic Assembly Near--infrared Deep Extragalactic Legacy
Survey (CANDELS). In particular, we select a sample of $\sim$107
massive ($M_*>10^{10}M_{\odot}$), passive ($SSFR<10^{-2} Gyr^{-1}$)
and morphologically spheroidal galaxies at $1.2<z<3$, taking advantage
of the panchromatic dataset available for GOODS, including VLT, CFHT,
Spitzer, Chandra and HST ACS+WFC3 data. We find that at $1<z<3$ the
passively evolving early--type galaxies are the reddest and most
massive objects in the Universe, and we prove that a correlation
between mass, morphology, color and star--formation activity is
already in place at that epoch. We measure a significant evolution in
the mass--size relation of passive early--type galaxies (ETGs) from
$z\sim3$ to $z\sim1$, with galaxies growing on average by a factor of
2 in size in a 3 Gyr timescale only. We witness also an increase in
the number density of passive ETGs of 50 times over the same time
interval. We find that the first ETGs to form at $z\gtrsim2$ are all
compact or ultra--compact, while normal sized ETGs (meaning ETGs with
sizes comparable to those of local counterparts of the same mass) are
the most common ETGs only at $z<1$. The increase of the average size
of ETGs at $0<z<1$ is primarily driven by the appearance of new large
ETGs rather than by the size increase of individual galaxies.

\end{abstract}

\section{Introduction}
Early--type galaxies have been the objects of many studies in the
recent years, as they contain crucial information about the evolution
of the galaxy content. First, they are by definition the oldest
objects at each epoch, and they can be considered relics of the star
formation activity that happened in the past; the age of the stellar
populations in these galaxies at $z\sim0$ are compatible with
redshifts of formations $z>2$ (Renzini~2006). Second, early--type
galaxies are the most massive objects in the local Universe,
containing the bulk of the stellar mass (Baldry~et~al.~2004).

Understanding the history of the assembly of these galaxies throughout
cosmic time is crucial to constrain models of galaxy evolution.  The
first generation of extremely massive ETGs ($M_{\odot}>10^{11}$) is
already in place at $z\sim2.5$ (Guo~et~al.~2012); their number density
dramatically increases during the $1<z<3$ epoch (Ilbert+10, Ilbert+13,
Cassata+11, Brammer+11), followed by a milder evolution at $0<z<1$
(Bell~et~al.~2004; Faber~et~al.~2007; Pozzetti~et~al.~2010). The
overall evolution of the ETGs depends strongly on stellar mass,
following a "downsizing" pattern: more massive ETGs build up
preferentially earlier than the less massive ones
(Arnouts~et~al.~2007; Marchesini~et~al.~2009; Ilbert~et~al.~2010,
2013; Cassata~et~al.~2011, Brammer~et~al.~2011).

At the same time, the massive passive ETGs are found to have 3--5
times smaller sizes at $z>1$ than in the local Universe
(Daddi~et~al.~2005; Trujillo et al. 2007; Toft et al. 2007; Zirm et
al. 2007; Cimatti~et~al.~2008; Van~Dokkum~et~al.~2008;
Buitrago~et~al.~2008), and thus they are 30--100 times denser. These
results ignited a debate about the possible mechanism producing the
size evolution: Naab,~Johansson~\&~Ostriker~(2009) claim that the
observed size evolution of passive ETGs can be explained by minor
merger events, while Khochfar~\&~Silk~(2006) propose that the observed
evolution can be explained by the variation of the amount of cold gas
available during the major merger events that produce ETGs: the most
massive ones formed at high--z, when major mergers were more gas rich
than at later epochs. Although Lopez--Sanjuan~et~al.~(2012) claimed
that the observed minor mergers at $0<z<1$ can account for up to 55\%
of the size growth of ETGs, Nipoti~et~al.~(2012) claimed that minor
and major mergers with spheroids are not sufficient to explain the
observed size evolution between $z\sim2.2$ and $z\sim1.3$.

From a theoretical point of view, some key questions on the formation
and evolution of ETGs are still open. We still do not know through
which mechanism the ETGs accrete their mass: is it through major
mergers at early epochs, as it is predicted by models of galaxy
formation (Shankar~et~al.~2010; Shankar~et~al.~2011) or is it through
the collapse of unstable disks, that have been shown to be numerous at
$z\sim2$ (Genzel~et~al.~2008; F\"{o}rster--Schreiber~et~al.~2009)? What
is the mechanism that shuts off the star formation in such objects,
turning them into passively evolving ETGs? The dependence of the
quenched ETG population on stellar mass and environment has been
recently elucidated on phenomenological grounds (Peng~et~al.~2010,
2012), but what remain to be understood are the concrete physical
mechanisms responsible for what is referred to as {\it mass quenching}
and {\it environment quenching} in these studies.


In this paper we take advantage of the wealth of data available in the
Chandra Deep Field South field, mainly gathered as a part of the GOODS
and CANDELS surveys, to select a robust sample of passive ETGs at
$1.2<z<3$, that we use to constrain the assembly of their mass content
as function of the size, complementing and completing the analysis
performed in Cassata~et~al.~(2011; C11 hereafter). In particular, we
dig into the passive population down to $M_*=10^{10} M_{\odot}$ up to
$z\sim3$, a mass regime between 5 and 10 times below $M^*$
(Ilbert~et~al.~2010).

Throughout the paper, we use a concordance cosmological model
($\Omega_{\Lambda}=0.7$, $\Omega_m=0.3$ and $H_0$=70 km s$^{-1}$
Mpc$^{-1}$, we assume a Salpeter initial mass function (IMF;
Salpeter~et~al.~1955) and we use AB magnitudes.

\section{Data and sample selection}
The GOODS--South field is the one of the best studied parts of the sky,
having been imaged with all the largest available telescopes (Hubble,
Spitzer, VLT, Chandra, XMM, Herschel, CFHT). The GOODS HST Treasury
Program (Giavalisco~et~al.~2004) provides ultradeep high--resolution
images in B--, V--, i-- and z--bands. Deep ground--based imaging in the
U--band is provided by VIMOS/VLT for the CDFS
(Nonino~et~al.~2009). Moreover, VLT/ISAAC imaged the CDFS in J--, H--
and K--bands.  Ultradeep Spitzer/IRAC imaging is also available in the
3.6, 4.5, 5.6 and 8.0 $\mu$m MIR channels and in the 24$\mu$m FIR
one. The field has also been recently imaged with Herschel/PACS at 100
and 160 $\mu$m as a part of the GOODS/Herschel program
(Elbaz~et~al.~2011).  About 3000 spectroscopic redshifts are also
available, among which 1200 are at $z>1$ (Cimatti~et~al.~2008;
Vanzella~et~al.~2008; Popesso~et~al.~2009; Kurk~et~al.~2013).

\begin{figure}
\begin{center}
\includegraphics[width=\columnwidth]{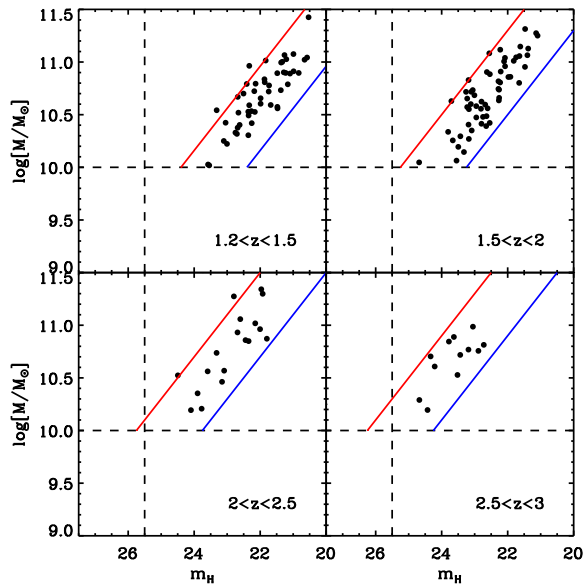}
\caption{Stellar mass vs. AB magnitude in the WFC3/F160W H--band filter
  in four redshift bins: $1.2<z<1.5$, $1.5<z<2$, $2<z<2.5$ and
  $2.5<z<3$. The vertical dashed line at $m_{F160W}=25.5$ indicates
  the magnitude at which the CANDELS H--band images are 80\% complete,
  the dashed horizontal line at $M/M_{\odot}=10^{10}$ shows the mass
  limit of our sample, the red and blue diagonal lines indicate, for
  each redshift bin, the stellar mass vs H--band magnitude relation for
  the reddest and bluest models available in our template
  grid. }\label{fig:completeness}
\end{center}
\end{figure}

The Cosmic Assembly Near--infrared Deep Extragalactic Legacy Survey
(CANDELS; Grogin~et~al.~2011; Koekemoer~et~al.~2011), the largest HST
campaign ever undertaken, is collecting high resolution images in the
F105W (Y--band), F125W (J--band) and F160W (H--band) filters for 5 of
the most intensively studied extragalactic fields: namely,
GOODS--South and North, EGS, UDS and COSMOS. In this paper we include
the first 4 epochs of CANDELS observations in GOODS--S, that cover the
central $\sim80$ arcmin$^2$ of the GOODS--S field, for a total
integration time of 2 orbits each in F125W (J--band) and F160W
(H--band). We include as well the WFC3 observations taken as a part of
the Early Release Science Program 2 (ERS2: GO 11359. PI: O'Connell;
Windhorst~et~al.~2011), that cover an additional $\sim40$ arcmin$^2$
area in the north part of the GOODS--S field, with integration times
of 2 orbits in each of the F098M (Y--band), F125W (J--band) and F160W
(H--band) filters. The final mosaics in Y--, J-- and H--band were
assembled using MultiDrizzle (Koekemoer et al. 2002), combining the
data to a 0.06'' pixel grid, and producing a PSF of $\sim$0.16'' in
our resulting WFC3 images. The $1\sigma$ fluctuations of the sky for
the ERS2 regions are 27.2, 26.6, and 26.3 AB $arcsec^{−2}$ in the Y, J
and H bands, respectively; for the 4--epoch CANDELS region, the
$1\sigma$ fluctuations are 26.6 AB $arcsec^{−2}$ for both the J-- and
H--bands.

We have built a multi--wavelength catalog using the TFIT procedures by
Laidler~et~al.~(2007), using the H--band as the detection image and
including the U--band from VIMOS, the B, V, i and z bands from
GOODS/ACS, the Y, J and H bands from CANDELS/WFC3 and the Ks band from
VLT/ISAAC (Guo~et~al.~2013, in~prep). This procedure allows us to match the
point--spread function of images having different resolutions and to
obtain homogeneous aperture magnitudes from the different images. The
catalog includes about 11,000 objects brighter than $m_H$=25.5, 4500
of which are at $1.2<z<3$, according to their spectroscopic or
photometric redshift (see below).

We included the spectroscopic redshifts available in literature for
the GOODS--S field (Vanzella~et~al.~2008; Cimatti~et~al.~2008;
Popesso~et~al.~2009; Kurk~et~al.~2013) for 2232 objects, and we
measured photometric redshifts for the remaining objects, using the
PEGASE 2.0 templates (Fioc~\&~Rocca--Volmerange~1997), following the
same procedure in Guo~et~al.~(2012).

We then fitted the SED from UV to 8 $\mu$m to the updated version of
Bruzual\&Charlot~(2003, CB07) in order to get accurate measurements of
stellar mass, E(B-V), age and SFR of the galaxies. In particular, we
use a Salpeter IMF (Salpeter~1955) with lower and upper masses of 0.1
and 100 $M_{\odot}$, we apply the Calzetti law (Calzetti~et~al.~2000)
to describe the dust extinction and we assume exponentially declining
star formation histories SFR(t)$\propto e^{(-t /\tau)}$, where $t$ is
{\it age} (i. e. the time passed since the peak of the star formation)
and $\tau$ is the characteristic time of the star formation event.
Maraston et al. (2010) showed that these exponentially declining SFR
models tend to overestimate the SFRs and underestimate stellar masses
for star--forming galaxies at z$\sim$2, while truncated exponentially
increasing models give better results; however, the effect of such
different SFH on passively evolving galaxies have not been conducted
yet, and we defer this analysis to a forthcoming paper. In any event,
these SED fits are only used to identify passively evolving galaxies.

We extracted a ``parent'' catalog of 1051 galaxies with
$M_*>10^{10}M_{\odot}$ and redshift $z>1.2$, from which we selected
only passive galaxies with Specific Star Formation Rate $SSFR<10^{-2}$
$Gyr^{-1}$ (332 galaxies). This SSFR limit is very restrictive:
galaxies with $SSFR=10^{-2}$ $Gyr^{-1}$ would need 100 Gyr to double
their stellar mass, if they continue to form stars at the present
rate.  We then eliminated all galaxies with a late-type morphology in
the H--band WFC3 image (based on visual inspection, similarly to C11)
and we also excluded galaxies detected in the Spitzer/MIPS 24$\mu$m
and Herschel/PACS 100$\mu$m channels, ensuring that the SFR is lower
than 10 $M_{\odot} yr^{-1}$ (Elbaz~et~al.~2011): the resulting sample
of passive ETGs contains 107 galaxies. For the remaining of the paper,
we define ``ETGs'' galaxies that are passive according to our criteria
(SSFR and no detection in Herschel/Spitzer) and have a spheroidal
morphology. 84\% and 92\% of the resulting 107 passive morphologically
spheroidal ETGs have sersic indices $n>2$ and $n>2.5$, respectively
(similarly to Bell~et~al.~2012): this ensures that our sample is not
contaminated by quiescent disks as in van~der~Wel~et~al.~(2011).

We used the GALFIT package (Peng~et~al.~2002) to fit the light profile
in the $H--$band (matching the rest--frame optical up to $z\sim3$):

\begin{equation}
I(r)=I_eexp\left\{-b_n\left[\left(\frac{r}{r_e}\right)^{1/n}-1\right]\right\},
\end{equation}

where $I(r)$ is the surface brightness measured at distance $r$, $I_e$
is the surface brightness at the effective radius $r_e$ and $b_n$ is a
parameter related to the S\'ersic index $n$. For $n$=1 and $n$=4 the
S\'ersic profile reduces respectively to an exponential and
deVaucouleurs profile. Bulge dominated objects typically have high $n$
values (e.g. $n>2$) and disk dominated objects have $n$ around
unity. Ravindranath~et~al.~(2006), Cimatti~et~al.~(2008),
Trujillo~et~al.~(2007) and van~der~Wel~et~al.~(2012) showed that
GALFIT yields unbiased estimates of the S\'ersic index and effective
radius for galaxies with S/N$>10$ and $r_e>0.03''$, independently of
the redshift of the source, thus demonstrating that the surface
brightness dimming is not an issue for this kind of study.

The PSF was obtained in each passband needed by averaging
well--exposed, unsaturated stars. We run GALFIT experimenting on
various sizes of the fitting region around each galaxy, and with the
sky either set to a pre--measured value or left as a free parameter. We
verified that the sizes and S\'ersic indices do not vary by more than
10\% in the various cases. The values that we show throughout the
paper were obtained with a free sky and $6\times6$ arcsec$^2$ fitting
regions. Any close-by object detected by SExtractor within each fitting
region was automatically masked out during the fitting procedure.

Figure~\ref{fig:completeness} illustrates the completeness of our
H--band selected catalog at different redshifts. In particular, we want
to assess our ability to recover ETGs, which are very red because of
the extremely constraining SSFR cut (SSFR$<10^{-2}$), as a function of
the stellar mass and redshift. The H--band image that we use in this
work is 80\% complete at least down to magnitude 25.5 for typical
$z\sim2$ spheroids with $R_e=0.125''$ (Guo~et~al.~2013, in~prep).  We
note that the 107 ETGs in our sample are detected at at more
than 10$\sigma$ in at least 4 filters blueward than the Balmer break,
ensuring good photometric redshifts and mass determination.

In Fig.~\ref{fig:completeness} we plot the stellar mass versus H--band
magnitude for the reddest models in our grid (thus the hardest to
detect), and we compare it with the distribution of our real ETGs at
$1.2<z<3$. As expected, all the real galaxies lie right of the reddest
model, meaning that, for a given stellar mass, they are brighter than
the model. Since the H--band imaging is 80\% complete down to
magnitude $m_H\sim25.5$ (Guo~et~al.~2013, in prep), we can conclude
that our sample is complete down to stellar mass $M_{\odot}=10^{10}$
up to $z\sim3$. This implies that the scarcity of objects with
$10^{10}M_{\odot}<M_*<10^{10.5}$ in the two highest redshift bins of
Fig.~\ref{fig:completeness} is a real effect, and it is not due to
mass incompleteness.

\section{The color--mass diagram at $1.2<z<3$}
\begin{figure}
\begin{center}
\includegraphics[width=\columnwidth]{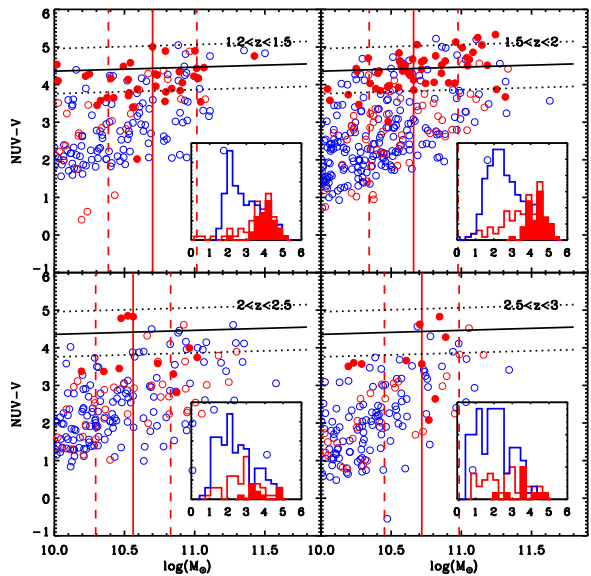}
\caption{The N$_{UV}$--V color vs mass in the optical rest--frame for
  in four redshift bins: $1.2<z<1.5$, $1.5<z<2$, $2<z<2.5$ and
  $2.5<z<3$. The filled circles, red empty circles and blue empty
  circles represent the passive early--type galaxies (spheroidal
  morphology and SSFR$<10^{-2}$), the morphologically early--type
  galaxies and the morphologically late--type galaxies,
  respectively. The black continuous and dashed lines show a fit to
  the ETGs in the first two bins combined. The insets show the color
  distribution in each bin, with the same color coding as the large
  figures. The red continuous and dotted lines show the median mass
  and scatter for the ETGs in the 4 redshift
  bins.}\label{fig:color_mass}
\end{center}
\end{figure}


In Figure~\ref{fig:color_mass} we show the N$_{UV}$ color--mass diagram
in four redshift bins ($1.2<z<1.5$, $1.5<z<2$, $2<z<2.5$ and
$2.5<z<3$) for the ``parent'' catalog (empty circles) and the ETGs
(filled red circles). The absolute magnitudes for the N$_{UV}$ and V
filters are computed by interpolating the observed photometry at the
location expected for the GALEX NUV 2500\AA~ and Johnson V 5500\AA,
respectively.  We separate the parent catalog in morphologically
early-- and late--types, based on the visual H--band morphology.

We observe a color bimodality up to $z\sim2$, with the red peak
dominated by ETGs and the blue distribution dominated by star forming
late--type galaxies. At $z>2$ the color distribution is unimodal, but
still the red galaxies are on average more massive than the blue
ones. In the two highest redshift bins, the bulk of objects displays
blue colors and only a ``tail'' of red objects is present.  Only about
50\% of the galaxies on the red sequence at such redshift are passive
according to our criteria.



We note that the ETGs are the reddest and most massive objects, and
that the red sequence is mostly populated by galaxies with
$M_*>10^{10.5}M_{\odot}$, at all redshifts. The average mass for ETGs
in the four redshift bins is around $M_*=10^{10.7}M_{\odot}$, with a
dispersion of 0.2 dex. This shows that the most massive ETGs are the
first to form at $z>1.2$, and that, as cosmic time goes by, new low
mass ETGs are formed. This result confirms earlier findings according
to which the most massive galaxies are the first to be quenched (e.g.,
Bundy~et~al.~2006; Franceschini~et al.~2006;
Cimatti,~Daddi~\&~Renzini~2006) and is in qualitative agreement with
the phenomenological model of Peng~et~al.~(2010).

We see that there is a clear correlation between the morphology, the
color and the star formation activity of the galaxies, at least up to
$z=2$: morphological early--type galaxies are typically red and
passive; morphological late--types are blue and star forming. This is
in good agreement with Whitaker~et~al.~(2011), who found a
red--sequence up to $z\sim3.5$ (but the authors did not analyze the
morphology of the galaxies on the red sequence), and
Bell~et~al.~(2012). These findings imply that the Hubble sequence, in
the sense of a correlation between stellar mass, star formation
properties and morphology, as observed in the local Universe, is
already in place at $z\sim3$. The mere existence of ETGs at $z\sim3$
indicates that such objects formed and were quenched at even higher
redshift; our estimate of the age is not accurate enough to accurately
age--date the bulk of the stellar component of such objects, but
presumably they accreted the bulk of their mass about 1 Gyr prior to
the observation, implying a formation redshift $z\sim5$ (similar to
Gobat~et~al.~2012).

\section{The mass--size relation for early--type galaxies at $1<z<3$}
\begin{figure}
\begin{center}
\includegraphics[width=\columnwidth]{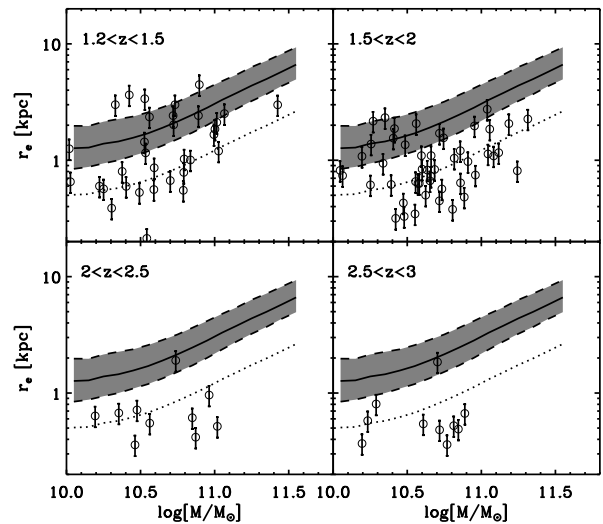}
\caption{The mass size relation in the optical rest--frame for in four
  redshift bins: $1.2<z<1.5$, $1.5<z<2$, $2<z<2.5$ and $2.5<z<3$. In
  all bins, the grey filled region indicates the locus occupied by
  SDSS passive galaxies at $0<z<0.1$: the continuous line shows the
  median of the distribution, and the dashed lines contain 68\% of the
  objects (from C11). Galaxies below the dotted line
  are defined as ultra--compact according to
  C11. }\label{fig:mass_size}
\end{center}
\end{figure}

In Figure~\ref{fig:mass_size} we show the mass--size relation for the
107 massive and ETGs at $1.2<z<3$, in four redshift bins. The
size is measured in the WFC3 H--band and matches the optical
rest-frame in the whole redshift range.The plot shows also the
mass--size relation for local ETGs drawn from the SDSS as derived by
C11, according to whom galaxies below the SDSS local relation (the
grey strip in Figure~\ref{fig:mass_size}) are defined as {\it compact}
ETGs (by definition, 17\% of local ETGs are {\it compact}), and those
more than 0.4 dex smaller than local counterparts of the same mass
(the dotted line in Figure~\ref{fig:mass_size}) are defined as {\it
  ultra--compact}.  According to these definitions, compact and
ultra--compact galaxies with a stellar mass $M_*=10^{10.5} M_{\odot}$
have respectively sizes smaller than about 1 and 0.5 kpc.  In our
sample, at $2.5<z<3$, $2<z<2.5$, $1.5<z<2$ and $1.2<z<1.5$,
respectively $\sim$90\% ($\sim$70\%), $\sim$90\% ($\sim$60\%),
$\sim$80\% ($\sim$50\%) and $\sim$70\% ($\sim$40\%) of the ETGs are
compact (ultra--compact) ETGs.

\begin{figure}[!t]
\begin{center}
\includegraphics[width=\columnwidth]{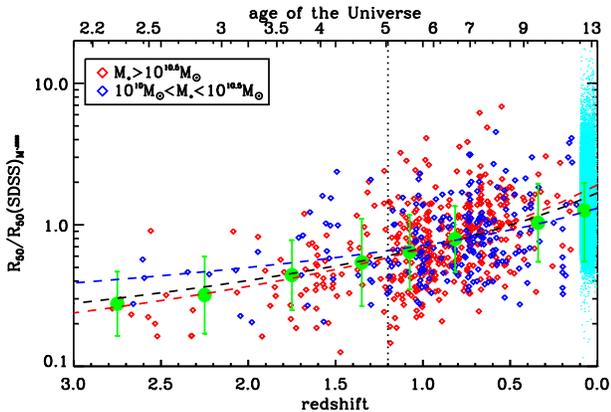}
\caption{Evolution of the size for passive galaxies at $0<z<3$,
  normalized to the SDSS. The blue and red points represent ETGs with
  $M_*<10^{10.5} M_{\odot}$ and with $M_*>10^{10.5} M_{\odot}$,
  respectively. Measures at $z<1.2$ (indicated by the dotted vertical
  line) come from C11. The green filled circles with
  their error bars represent the median size and scatter in bin of
  redshift. The black, blue and red dashed line show the best fit for
  all the ETGs, the ETGs with $M_*<10^{10.5} M_{\odot}$ and ETGs with
  $M_*>10^{10.5} M_{\odot}$, respectively.}\label{size_redshift}
\end{center}
\end{figure}
In Figure~\ref{size_redshift} we report the size as a function of the
redshift for our sample of 107 massive ETGs at $1.2<z<3$, together
with data at $0<z<1.2$ from C11: measurements for ETGs at $0.3<z<1.2$
are done in the $z$--band, that matches the optical rest--frame (see
C11 for details); ETGs in the local universe ($0<z<0.1$) are drawn
from the SDSS, and the measurements of their size are taken by the DR7
NYU Value--Added Galaxy Catalog (Blanton~et~al.~2005, see C11 for
details).  Galaxy sizes are normalized to the average size of the SDSS
ETGs with the same stellar mass. We also tried to normalize the size
to the best--fit $R_e\propto M_{\odot}^{0.55}$ relation for each
redshift (as for example in Cimatti,~Nipoti~\&~Cassata~2012), but the
essence of the results discussed here did not change. In the same
figure, we report the median normalized size for 8 redshifts, along
with the scatter of the distribution: the $1-\sigma$ standard
deviation of the distribution is about 0.25 dex at all redshifts. We
parameterize the evolution of the average size as
$\left<r_e\right>\propto~(1+z)^{\alpha}$, and we fitted the global
population of ETGs, as well as ETGs with $M\lessgtr M^{10.5}M_{\odot}$
separately. To avoid the fit being completely driven by the
$\sim$100,000 SDSS galaxies at $0<z<0.1$, we exclude them from the
fit. We find $\alpha=-1.29\pm0.10$ for the global population of ETGs,
$\alpha=-0.89\pm0.16$ for ETGs with $M_*<M^{10.5}M_{\odot}$ and
$\alpha=-1.50\pm0.12$ for ETGs with $M_*>M^{10.5}M_{\odot}$. This
means that the size evolution is faster for high mass ETGs than for
the low mass ones, in qualitative agreement with Ryan~et~al.~(2012).
Interestingly, this result does not change if only the C11 galaxies
with $z<1.2$ are included in the fit: we find $\alpha=-1.18\pm0.15$,
$\alpha=-0.90\pm0.22$ and $\alpha=-1.33\pm0.18$ for all the ETGs, the
ETGs with $M_*<M^{10.5}M_{\odot}$ and the ETGs with
$M_*>M^{10.5}M_{\odot}$, respectively.

These values are in good agreement with results by
Cimatti,~Nipoti~\&~Cassata~(2012), who found $-1.25<\alpha<-0.8$, and
slightly smaller (in absolute value) than the value published by
Damjanov~et~al.~(2011), who found $\alpha=-1.62$. However, we stress
that the sample used here (the 107 ETGs at $z>1.2$ plus the ETGs at
$z<1.2$ by C11 is complete in mass down to $M_*=10^{10}
M_{\odot}$ up to $z=3$, and it selected according to the same criteria
at all redshift, thus being more homogeneous and complete than the
ones used in these works, which are compilations of different samples
published in literature.

\section{Evolution of the number density in passive galaxies at $1<z<3$}

\begin{figure}
\begin{center}
\includegraphics[width=\columnwidth]{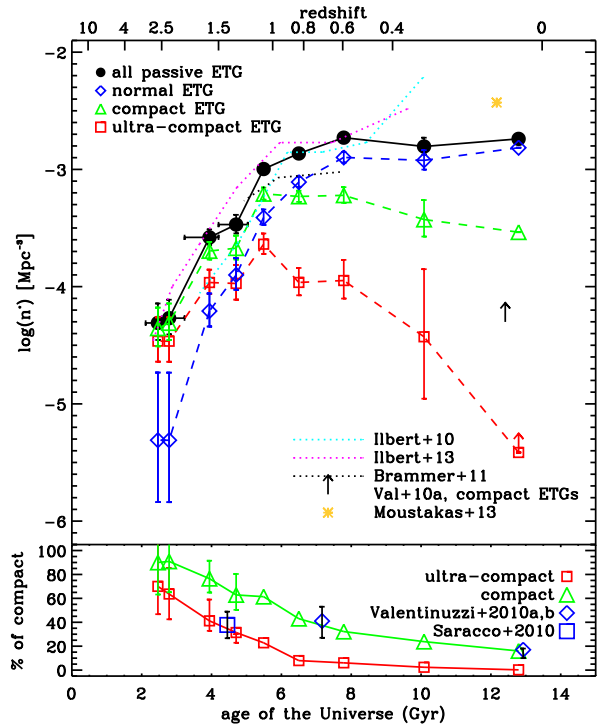}
\caption{{\it Top panel:} number density as a function of the age of
  the universe and redshift of all ETGs (black circles),
  normal size ETGs (blue diamonds), compact ETGs (green triangles) and
  ultra--compact ETGs (red squares). As a comparison, we report the
  data for passively evolving galaxies from Ilbert~et~al.~(2010, cyan
  dotted line), Ilbert~et~al.~(2013, purple dotted line),
  Brammer~et~al.~(2011, black dotted line) and Moustakas~et~al.~(2013,
  yellow filled star). We also report the lower limit estimate of the
  number density of compact ETGs by Valentinuzzi~et~al.~(2010a, black
  upward arrow), fully compatible with our $z\sim0$ estimate based on
  SDSS. {\it Bottom panel:} Fraction of passive galaxies that are also
  compact (green triangles) and ultra--compact (red squares), as a
  function of the age of the universe or the redshift.  }\label{fig5}
\end{center}
\end{figure}

In Section~4 we saw that the mass--size relation of massive ETGs
evolves fast with the redshift between $z\sim3$ and $z\sim1.2$: at
$z>2$ about 90\% of the ETGs are compact (i.e. $1-\sigma$ below the
local relation), with that fraction dropping to 70\% at
$z\sim1.2$. From Fig~4 we also learned that the average size of ETGs
roughly doubles over the same redshift interval. The evolution of the
average size of galaxies illustrated in in Figure~4 can be driven by
two simultaneous processes: the size increase of each galaxy, possibly
due to (minor) dry mergers (e.g., Naab~et~al.~2007;
Nipoti~et~al.~2012; Oser~et~al.~2012) and to the appearance of new
ETGs with a size distribution progressively shifted to larger sizes
(Valentinuzzi~et~al.~2010a,b; Cassata~et~al.~2011; Newman~et~al.~2012;
Poggianti~et~al.~2012; Carollo~et~al.~2013). We know that the number
density of ETGs dramatically increases with cosmic time
(i.e. Ilbert~et~al.~2010; Pozzetti~et~al.~2010), hence the size
evolution can not be entirely attributed to the former mechanism, as
it has been recently attempted by Oser~et~al.~(2012). In order to try
to constrain the relative importance of the two mechanisms, in this
section we study the number density of ETGs of different physical
sizes as a function of the redshift and the cosmic time.

In Figure~\ref{fig5} we report the evolution of the number density of
ETGs between $z=3$ and $z=1.2$ from this work, together with the same
values at $0<z<1.2$ taken from C11. The only difference with C11 is
that here we rebinned the data at $0<z<1.2$ in a slightly different
manner, to better sample the range $0.1<z<0.7$, that covers about half
of the age of the Universe and was sampled with only one bin in
C11. We stress that the uniqueness of this work is that we exploit for
the first time a sample of ETGs complete down to $M_*=10^{10}
M_{\odot}$, extending the analysis of C11 up to $z\sim3$, when the
Universe was just 2 Gyr old. We find that the number density of ETGs
increases by a factor of 50 in a time span of only 3.5 Gyrs, between
$z\sim3$ and $z\sim1$. Thus, the epoch $1<z<3$ is critical for the
build--up of the number of massive ETGs, in good agreement with e.g.,
Ilbert~et~al.~(2010, 2013) and with the phenomenological model of
Peng~et~al.~(2010). The number density of ETGs then further doubles in
the 10 Gyr time span between $z\sim1$ and $z\sim0$, which is primarily
due to the increase of low--mass ETGs (e.g., Ilbert~et~al.~2010;
Franceschini~et~al.~2006; Cimatti,~Daddi~\&~Renzini~2006;
Peng~et~al.~2010). Our measurements for the number density evolution
of the global population of ETGs is in good agreement with
Ilbert~et~al.~(2010, 2013) and Brammer~et~al.~(2011), obtained with
much larger and less cosmic variance prone samples.

We see that the first ETGs to appear at $z\sim3$ are virtually all
compact or ultra--compact. The very rapid increase in the number
density of the global population of ETG between $z\sim3$ and $z\sim1$
is caused by the fast and steady increase of the number densities of
all normal, compact, and ultra--compact ETGs. By $z\sim1.5$, the
normal ETGs become more numerous than the ultra--compact, and at
$z\sim1$ they are the most common ETGs, exceeding the number of
compact ETGs. 

After $z\sim1$, the number density of normal ETGs keeps steadily
increasing, albeit at a modest rate. At the same time, the evolution
of compact and ultra--compact ETGs decouples from that of the large
ETGs, with their number density steadily decreasing. In the local
Universe, the compact galaxies are (by definition, as they are defined
to be 1-$\sigma$ below the local relation) 17\% of the global
population of ETGs, and the ultra--compact formally disappear almost
completely by $z\sim0$. However, it has been argued that the SDSS
database may be biased against very compact galaxies
(Scranton~et~al.~2002; Shih~\&~Stockton~2011;
Valentinuzzi~et~al.~2010a; Carollo~et~al.~2013).  We note that our
estimate of the number density of compact ETGs at $z\sim0$ is $\sim$5
times higher than the lower limit determined by
Valentinuzzi~et~al.~(2010a), thus formally in agreement, and that our
new point at $z\sim0.4$ seems to confirm the mild decrease. From
$z\sim1$ to $z\sim0$ the number density of compact ETGs decreases by a
factor of around 2. The picture is less clear for the ultra-compact
ETGs: the SDSS imaging does not offer enough spatial resolution to
identify ultra-compact ETGs, and thus we indicate the number density
of ultra-compact ETGs as a lower limit in Figure~5. The new point at
$z\sim0.4$ seems to support the fast decrease of the the number
density of ultra-compact ETGs at $0<z<1$, but the large error bar does
not help to unambiguously fix the issue. In the end, the evolution
seems to be strongly size--dependent, with a faster decrease for
smaller ETGs, even though we can not draw any strong conclusion due to
the $z\sim0$ uncertainties. This differential evolution is in quite
good agreement with recent results by Carollo~et~al.~(2013), who found
a even milder decrease of the number density of ETGs with sizes
smaller than 2 kpc (that is slightly larger than our ``compactness''
definition: according to our definition, ETGs of
$M_*=10^{10.5}M_{\odot}$ are compact if their size is smaller than 1
kpc).

\section{Discussion and conclusions}
In this paper we analyzed a sample of passive (SSFR$<10^{-2}Gyr^{-1}$)
and massive ($M_*>10^{10} M_{\odot}$) ETGs at $1.2<z<3$. The deep
H--band imaging ensures that the sample is complete down to
$M_*=10^{10} M_{\odot}$, about an order of magnitude below $M^*$
(Ilbert~et~al.~2010) and allows us to study the morphology of our
galaxies in the optical rest--frame up to $z\sim3$. Moreover, the
dense multiwavelength coverage ensures robust redshift measurements
(for the objects without a spectroscopic redshift), accurate mass and
star formation activity determinations. The Herschel and Spitzer
ancillary data allowed us to reinforce the passivity selection based
on the SED fitting procedure.

The key finding of this paper are: 1. the identification up to
$z\sim3$ of passive and massive ETGs, that dominate the red sequence
(well defined up to $z\sim2$, less clear at $2<z<3$, see
Figure~\ref{fig:color_mass}); 2. the accurate determination, with the
evolution of average size for ETGs at $0<z<3$, reported in
Figure~\ref{size_redshift}; 3. the robust determination of the
differential evolution of the spatial abundance of massive,
i.e. stellar mass $M_*>10^{10}$ M$_{\odot}$, passive galaxies as a
function of redshift and as a function of the their size, a proxy of
their average projected central stellar density, shown in
Figure~\ref{fig5}.

At high redshift, i.e. $z\sim3$, the size distribution of ETGs is
heavily tilted toward small sizes, and massive and passive galaxies
are predominantly high--stellar density systems. This implies that the
mechanism through which these galaxies accrete their mass and
eventually become passive at those early epochs leaves a remnant that
is very compact (Cimatti~et~al.~2008). Many different mechanisms
leading to the formation of compact remnants have been proposed in
literature: Dekel~et~al.~(2009) argued that disk instabilities lead to
the formation of compact remnants; Hopkins~et~al.~(2008),
Wuyts~et~al.~(2010) and Bournaud~et~al.~(2011) showed that gas--rich
mergers can form compact galaxies; Johansson,~Naab,~\&~Ostriker~(2012)
proposed that compact galaxies can formed by in situ star formation
driven by cold flows. 

By $z\sim 1$ the abundance of small ETGs reaches its peak, followed by
a decline, apparently faster for the ultra--compact than for the
compact ones (see Section~5 and discussion of the uncertainties for
$z<0.5$). Apparently, whatever set of mechanisms was producing
ultra--compact passive galaxies at $1\lesssim z\lesssim 2.5$ is not
working anymore at later epochs. Moreover, the size distribution of
the new ETGs that become passive between $z\sim1$ and $z\sim0$ is
gradually moving towards larger sizes, implying that the mechanism
through which star formation is quenched in these massive ETGs acts in
such a way as to leave larger quenched remnants as (cosmic) time goes
by, a tendency which accelerates at $z<1$.  This may be a natural
consequence of the {\it inside-out} growth of the disks (a widely
entertained notion, Samland~\&~Gerhard~2003; Brook~et~al.~2006;
Mu\~noz--Mateos~et~al.~2011), with such disks being the likely
precursors to passive ETGs. So, bigger disks would leave bigger ETG
{\it remnants}.



In the $0<z<1$ epoch the average size of the ETGs steadily increases
(see Figure~4). This evolution might be due to two different
mechanisms (or a mix of the two): on one hand, the individual ETGs
increase in size, by minor merging or accretion; on the other hand new
already large ETGs, that never passed through the compact or
ultra--compact phase, appear. In this respect, the evolution of the
number densities of ETGs of different sizes, constrained in Figure 5,
can help to weigh the relative importance of two mechanisms. In
particular, even assuming that all the disappearing compact and
ultra--compact ETGs are transformed in ETGs of normal size and
density, in a cascade of merging and interactions to form
normal--sized galaxies (e.g. Huang et al. 2012), Figure~5 shows that
they are not numerous enough to support the increase of the total
number of normal ETGs observed over the same time interval. In fact,
between $z\sim1$ and $z\sim0$ the number density all ETGs increases by
about $\sim0.8\times10^{-3} Mpc^{-3}$, while the number density of
compact ETGs that disappear is only
$\sim0.3\times10^{-3}Mpc^{-3}$. This implies that the observed rate at
which compact galaxies disappear can not sustain the observed increase
of the number of normal ETGs; thus the growth of the individual ETGs
is not the dominant mechanism causing the average size of ETGs to
increase (Fig.~\ref{size_redshift}), in agreement with
Carollo~et~al.~(2013).

At the same time, the decrease of the number density of compact and
ultra--compact ETGs between $z\sim1$ and $z\sim0$ indicates that a
fraction of these galaxies either become active, because of a sudden
refurbishment of fresh gas, or --more likely-- they grow in size, via
minor merging or smooth accretion, becoming normal-sized
ETGs. 



Before concluding, we also wish to comment on the cosmic rise and fall
of the ultra--compact passive galaxies that this paper has
documented. The extreme compactness and large stellar mass of these
galaxies argue against hierarchical merging of stellar systems as the
primary mechanism for the assembly of their stellar mass (Giavalisco
et al. in prep.). Rather, it is likely that they emerged in their
final state after the quenching of star--forming galaxies of the same
size and density. In this case the rise and fall of ultra--compact
ETGs are the result of four different processes: 1) the rate at which
ultra--compact star--forming galaxies form in this mass range (roughly
$M_*>10^{10}$ M$_{\odot}$) ; 2) the rate at which they are destroyed
before they quench, i.e. they are transformed from ultra--compact into
galaxies of lower density by merging and interactions; 3) the rate at
which the ultra--compact star--forming galaxies shut down star
formation and become passive; 4) the rate at which the ultra--compact
passive galaxies are transformed into galaxies of lower stellar
density by merging and interactions. 1) and 2), although not directly
constrained by this work, define the ``reservoir'' of compact
star--forming galaxies from which compact ETGs can form; the effects
of mechanism 3) and 4) in shaping the number density evolution of
compact ETGs has been extensively discussed in this paper.

Interestingly, Wuyts~et~al.~(2011) and Barro~et~al.~(2013) identified
a rich population of compact star forming galaxies at $1.5<z<3$, whose
number densities and physical properties are compatible with being the
progenitors of the compact and ultra--compact ETGs. The authors
speculate that those compact star forming galaxies are formed from a
gas--rich process (merger or disk instabilities) that at first induce
a compact starburst and then feed an AGN, which quenches the star
formation turning these objects into compact ETGs. They find that
these compact star forming objects are very rare at $z<1.5$,
supporting our conjecture that after that epoch the mechanism through
which ETGs are formed produce a large remnant. However, the definition
of ETGs in Barro~et~al.~(2013), as well as the definition of which
star forming galaxies can transform in ETGs differ from ours. So, we
looked for progenitors of our ETGs in a companion paper,
Williams~et~al.~(2013, in preparation): we have identified a sample of
compact star forming galaxies which may be progenitors of our
high--redshift passive sample, (which represent some of the first ETGs
to appear in the universe). This sample have consistent SFRs and
stellar masses with our ETGs, and are evolutionarily consistent
assuming physically motivated SFHs. This sample demonstrates that
massive and active enough compact SF galaxies exist at $z>3$ to
account for the number density of compact ETGs presented here,
assuming mass buildup by purely in--situ star--formation, further
justifying the evolutionary link between compact SF and compact
passive galaxies.

To summarize, our main findings are:
\begin{itemize}
\item We find that at $1<z<3$ the passively evolving ETGs are the
  reddest and most massive objects in the Universe. This implies that
  an embryo of the Hubble Sequence, in the sense of a correlation
  between morphology, mass, color and star--formation activity of
  galaxies, is already in place at $z\sim3$. We observe a scarcity of
  ETGs with $M_*<10^{10.5} M_{\odot}$, with the majority of our ETGs
  having $M_*>10^{10.5} M_{\odot}$. Since we accurately set our mass
  completeness to $M_*=10^{10} M_{\odot}$, we can conclude that that
  scarcity is not due to an observational bias. Hence, at $z>1.2$ the
  mechanism producing ETGs leaves a remnant that is preferentially
  very massive. This result reinforces previous claims for a
  ``downsizing'' pattern of the mass assembly of ETGs. A possible
  interpretation is that at that early epoch the process that suddenly
  quenches the star--formation activity in some objects, transforming
  them into passively evolving ETGs, is effective preferentially for
  objects with stellar mass above $M_*=10^{10.5} M_{\odot}$. This
  result is in qualitative agreement with Peng~et~al.~(2010), who
  highlighted the prominent role of mass quenching at high redshift.
\item We measure a significant evolution of the mass--size relation of
  ETGs from $z\sim3$ to $z\sim1$, with the average size of galaxies
  increasing by roughly a factor of $\sim$2 over this redshift
  interval, corresponding to 3 Gyrs of cosmic time. The evolution of
  the size of ETGs is faster for galaxies with $M_*>10^{10.5}
  M_{\odot}$ than for those with $M_*<10^{10.5} M_{\odot}$. About 90\%
  (70\%) of the ETGs at $z>2$ are compact (ultra--compact). We find
  that the average size of ETGs between $z\sim3$ and $z\sim0$ evolves
  with the redshift following a simple power law
  $r_e\sim(1+z)^{\alpha}$, with $\alpha=-1.18\pm0.15$.  If ETGs with
  $M\lessgtr M^{10.5}M_{\odot}$ are fitted separately, we find a
  marginally steeper power $\alpha$ for the most massive ETGs
  ($\alpha=-1.33\pm0.18$), indicating a faster size evolution.
\item We witness the build up of the most massive ETGs, with their
  number density increasing by 50 times between $z\sim3$ and
  $z\sim1$. We find that $90$\% of ETGs at $z>2$ are compact or
  ultra--compact, indicating that the event through which such first
  ETGs accrete their mass leaves a remnant that is very compact. As
  the cosmic time goes by, new ETGs tend to be larger and larger, with
  the ``normal'' sized ETGs (meaning those objects with sizes
  comparable to the local ETGs of the same mass) becoming the most
  common ETGs at $z\sim1$. At $z>1$ the mechanisms creating new
  ultra--compact ETGs prevail on those destroying them (merging,
  smooth accretion), resulting in a net increase of their number
  density; at $z<1$ the balance between such mechanisms inverts, with
  the processes decreasing the stellar density of ultra--compact
  galaxies finally prevailing.  Thus, the number density of compact
  passive ETGs starts decreasing, although the measure of such
  decrease to $z=0$ remains quite uncertain. Still, such decrease
  would not account for the increased number of ``normal--size"
  galaxies, even if all the compact, passive ETGs at $z=1$ were to
  disappear by $z=0$, Therefore, the evolution of the average size of
  ETGs at $0<z<1$ is mainly due to the appearance of newly quenched
  ETGs that are born large, rather than to the size increase of
  individual galaxies.
\end{itemize}

\begin{acknowledgements} 
Paolo Cassata acknowledges support from ERC advanced grant
ERC-2010-AdG-268107-EARLY.
\end{acknowledgements}

\end{document}